\documentclass[aps,prl,twocolumn,groupedaddress,superscriptaddress,showpacs]{revtex4}

\usepackage{amsfonts,amssymb,amsbsy,amsmath}
\usepackage{graphicx,txfonts}

\newcommand{\avrg}[1]{\left\langle #1 \right\rangle}

\begin{document}

\title{Temporal Network Sparsity and the Slowing Down of Spreading}

\author{Juan Ignacio Perotti}
\affiliation{BECS, School of Science, Aalto University, P.O. Box 12200, FI-00076, Finland}

\author{Hang-Hyun Jo}
\email[E-mail: ]{johanghyun@postech.ac.kr}
\affiliation{BECS, School of Science, Aalto University, P.O. Box 12200, FI-00076, Finland}
\affiliation{BK21plus Physics Division and Department of Physics, Pohang University of Science and Technology, Pohang 790-784, Republic of Korea}

\author{Petter Holme}
\affiliation{Department of Energy Science, Sungkyunkwan University, 440-746 Suwon, Republic of Korea}
\affiliation{Department of Physics, Ume{\aa} University, 90187 Ume\aa, Sweden}
\affiliation{Department of Sociology, Stockholm University, 10961 Stockholm, Sweden}

\author{Jari Saram\"aki}
\affiliation{BECS, School of Science, Aalto University, P.O. Box 12200, FI-00076, Finland}


\begin{abstract}
Interactions in time-varying complex systems are often very heterogeneous at the topological level (who interacts with whom) and at the temporal level (when interactions occur and how often). While it is known that temporal heterogeneities often have strong effects on dynamical processes, e.g.\ the burstiness of contact sequences is associated with slower spreading dynamics, the picture is far from complete. In this paper, we show that temporal heterogeneities result in \emph{temporal sparsity} at the time scale of average inter-event times, and that temporal sparsity determines the amount of slowdown of  Susceptible-Infectious (SI) spreading dynamics on temporal networks. This result is based on the analysis of several empirical temporal network data sets. An approximate solution for a simple network model confirms the association between temporal sparsity and slowdown of SI spreading dynamics. Since deterministic SI spreading always follows the fastest temporal paths, our results generalize---paths are slower to traverse because of temporal sparsity, and therefore all dynamical processes are slower as well. 
\end{abstract}


\maketitle

Complex systems usually exhibit strongly heterogeneous interaction patterns between their components. 
This is evident in the topology of networks that represent how the components are interconnected. Empirical complex 
 networks are usually neither very regular nor purely random.
 Rather, they display heterogeneities at multiple scales, from the broad distributions of node degrees~\cite{barabasi1999emergence} and interaction strengths~\cite{barrat2004architecture} to 
community structure~\cite{fortunato2010community}.
 
  In addition to topology, heterogeneities are abundant in time---when timings of interactions in networks are investigated in detail with the temporal network approach~\cite{holme2012temporal}, it is  seen that their inter-event times are also broadly distributed, i.e.\ event trains are bursty~\cite{neuts1993burstiness,paxson1995wide,masoliver2003continuous,kepecs2003information,vazquez2007impact}. 
Much of the research on timings of interactions in temporal networks has focused on this burstiness, in particular on its effects on the speed of spreading processes~\cite{vazquez2007impact,iribarren2009impact,Miritello2011,small2011karsai,kivela2012multiscale,jo2014analytically,scholtes2014causality,Horvath2014}. Typically, spreading processes are slower on networks with bursty contact sequences (with some exceptions~\cite{Horvath2014}). 
Additionally, there are other types of  temporal heterogeneities on multiple time scales---from variation in the activity levels of nodes and links~\cite{jo2012circadian} to their finite lifespans~\cite{holme2014birth}, and to correlated event sequences forming temporal motifs~\cite{kovanen2013temporal}. 

In this paper, we show that the general existence of  temporal heterogeneities results in \emph{temporal sparsity} of networks, limiting the effective number of links that are active at any point in time. As a consequence,  the level of sparsity observed in empirical temporal networks determines the relative speed of spreading processes compared to temporally homogeneous networks.

\begin{figure}
  \includegraphics[width=\columnwidth]{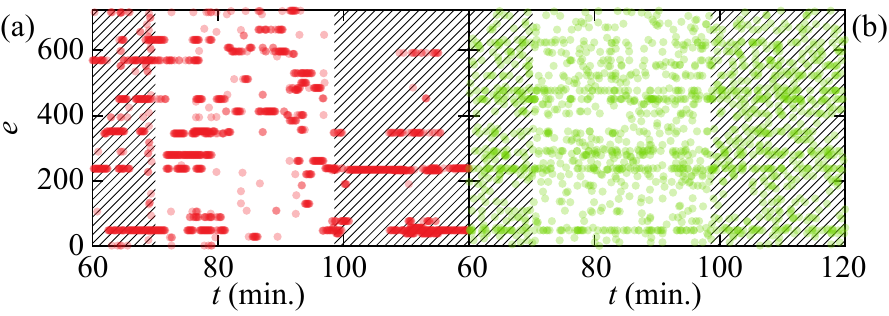}
  \caption{
(Color online).
Link activity profiles for the {\em Reality} temporal network (a) and for the corresponding Uniform Times reference model (b).
Each point represents an event involving link $e$ at time $t$.
For clarity, we show only 60 mins.\ of activity.
The unshaded bands represent time windows corresponding to the average inter-event time, $\avrg{\tau}=28.5$ mins.
Altogether 97 links are active within the window in (a), while in (b) the corresponding number is 313.}
\label{fig:1}
\end{figure}

Formally, temporal networks are sets of nodes $v \in \mathcal{V}$ and events $s \in \mathcal{E}$, where each event is a triple $s=(v,u,t)$ that denotes a contact  between nodes $v$ and $u$ at time $t \in [0,T]$.
Here, $[0,T]$ is the finite period of observation; in this paper, we apply periodic temporal boundary conditions to event sequences. Further, we assume that events are instantaneous and undirected. We use several empirical data sets on temporal networks, depicting timed email communications ({\em E-mail 1} and {\em E-mail 2}), physical proximity ({\em F2F}, {\em Hospital}, and {\em{Reality}}), interactions in online social networks and internet communities ({\em FB}, {\em Forum}, {\em Messages}, and {\em Dating}), and mobile phone calls in an European city ({\em MPC}). For references and details of the networks, see Table~\ref{table:table1}. 
We only consider nodes and events that belong to the largest connected component (LCC) of the corresponding networks aggregated over $[0,T]$. Furthermore, early/late transitory periods where networks grow or shrink are filtered out from some data sets. 

\begin{table}
\begin{ruledtabular}
\begin{tabular}{l|rrrrrr}
Name  & $N$ & $M$ & $E$ & $T$ & $\delta T$ & $k_\mathrm{eff}^{H}$ \\
\hline
{\em E-mail 1}~\cite{ebel2002scale} & 56,576 & 92,013 & 431,138 & 112 d & 1 s & 0.23 \\
{\em E-mail 2}~\cite{eckmann2004entropy} & 3,186 & 31,856 & 308,726 & 82 d & 1 s & 2.82 \\
{\em F2F}~\cite{isella2011s} & 410 & 2,765 & 17,298 & 8 h & 20 s & 0.99 \\
{\em FB}*~\cite{viswanath2009evolution} & 31,359 & 120,229 & 566,305 & 15,000 h & 1 s & 1.24 \\
{\em Forum}*~\cite{karimi2014structural} & 6,625 & 129,667 & 1,359,075 & 2,400 d & 1 s & 1.21 \\
{\em Hospital}~\cite{vanhems2013estimating} & 75 & 1,139 & 32,424 & 4 d & 20 s & 0.69 \\
{\em Messages}*~\cite{karimi2014structural} & 22,695 & 56,929 & 280,717 & 3 d & 1 s & 0.32 \\
{\em MPC}~\cite{pan2011path} & 10,448 & 15,506 & 601,116 & 120 d & 1 s & 0.63 \\
{\em Dating}*~\cite{holme2004structure} & 17,009 & 50,124 & 185,578 & 250 d & 1 s & 0.91 \\
{\em Reality}~\cite{PhysRevLett.110.198701} & 64 & 722 & 13,131 & 8.6 h & 5 s & 0.56 
\end{tabular}
\end{ruledtabular}
\caption{
Details on the used temporal network data sets.
$N$ denotes the number of nodes, $M$ the number of links, $E$ the number of events, $T$ the sampling time of the original data set, $\delta T$ the time resolution, and $k_\mathrm{eff}^{H}$ is the effective degree. Transient periods are removed from networks marked with an asterisk.
}\label{table:table1}
\end{table}

We begin by illustrating some typical heterogeneities in temporal networks.
Fig.~\ref{fig:1}(a) shows the activity profile of the links in the {\em Reality} temporal network
for 60 mins.~of activity. In the profile, each point represents an interaction event taking place at time $t$ on link $e$.
Clearly, activity is at different times concentrated on different groups of links while most links are inactive.
For reference, Fig.~\ref{fig:1}(b) shows a version of the same data where real event times have been replaced by 
times picked uniformly at random from $[0,T]$, i.e., using the Uniform Times (UT) reference model. In the UT
model, the number of events and the average inter-event times on each link are the same as in the original data, while the 
sequences are otherwise random. It is worth noting that in Fig.~\ref{fig:1}(b), the events are not only more homogeneously distributed across time, but also across links. 

In order to quantify the heterogeneity of the activity level across links in a given time window, we adopt a notion from statistical physics---the  multiplicity, or number of microscopic configurations associated with the macro-state of a system~\cite{kittel1980thermal}. More specifically, given the entropy, $H_e=-\sum_e p_e\ln p_e$, where $p_e$ is the fraction of events that pertains to link $e$  in the time window, we introduce an entropy-based \emph{effective number of links}:
\begin{equation}
\label{eq:1}
M_\mathrm{eff}^{H}=\exp(\avrg{H_e}).
\end{equation}
The average $\avrg{H_e}$ is taken by sampling over time windows; we  use the average inter-event time of links $\avrg{\tau}$ as the time window length. $M_\mathrm{eff}^{H}$ equals the number of active links in the time window if the number of events per link is constant; however, for broadly distributed event numbers, $M_\mathrm{eff}^{H}$ is significantly smaller.

To put to the values of $M_\mathrm{eff}^{H}$ on a scale, we introduce the {\em temporal sparsity coefficient},
\begin{equation}
\label{eq:2}
\mu^{H}=M_\mathrm{eff}^{H}/M_{\mathrm{eff},\mathrm{UT}}^{H}\in (0,1],
\end{equation}
where $M_{\mathrm{eff},\mathrm{UT}}^{H}$ is the effective number of links for the UT reference model with events uniformly distributed in time.
If $\mu^{H}$ is close to unity, the level of temporal heterogeneities in the network is very small. The smaller $\mu^{H}$, the 
more severe the heterogeneities, and the more temporally sparse the network.

We next illustrate the power of this concept by measuring how temporal sparsity affects SI spreading dynamics. We use our empirical data sets as substrates, where simulated spreading processes start at random times and at random nodes set to the infectious state while the rest are susceptible. In the spreading dynamics, an event connecting an infectious and susceptible node always results in infecting the latter. We apply this rule to the empirical event sequence until half of the nodes are infected, and measure the time it takes to arrive at this point, $t_{1/2}$, and average over numerous runs. For reference, we compute $\avrg{t_{1/2}}_{\mathrm{UT}}$, the average time it takes to infect half of the network for UT model versions of the empirical sequence, and then determine the {\em slowdown coefficient}
\begin{equation}
\label{eq:3}
\eta=\avrg{t_{1/2}}_{\mathrm{UT}}/\avrg{t_{1/2}} \in(0,1].
\end{equation}
The smaller $\eta$ is, the slower the spreading processes are in the original temporal network as compared to the UT ensemble.
 
\begin{figure}
  \includegraphics[width=.65\columnwidth]{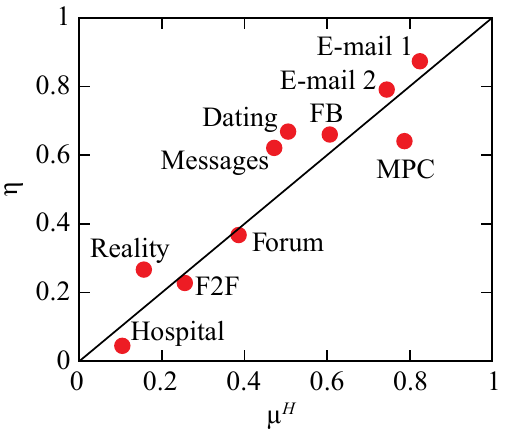}
  \caption{
(Color online). The dependence of the slowdown coefficient $\eta$  on the temporal sparsity coefficient $\mu^{H}$ for the different temporal network data sets. $\eta$ is seen to follow $\mu^{H}$ almost linearly, with $R^2=0.86$ between the data points and identity function (solid line).
}
\label{fig:2}
\end{figure}

 Figure~\ref{fig:2} displays the slowdown coefficient $\eta$ as a function of the sparsity coefficient $\mu^{H}$ for our empirical temporal networks. There is an almost linear dependence,  $\eta \simeq \mu^{H}$, indicating that the more temporally sparse the networks are because of  heterogeneities, the slower the SI spreading process is.  In other words, when temporal heterogeneities limit the effective number of links, waiting times at nodes are increased since there are less available links to carry the infection forward.

\begin{figure}
  \includegraphics[width=0.8\columnwidth]{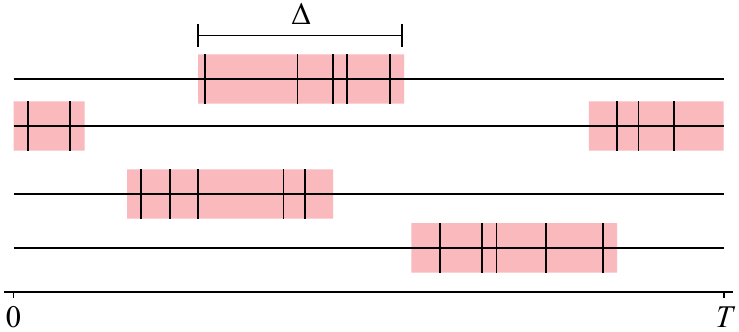}
  \caption{
(Color online).
Illustration of the Single-Burst (SB) temporal network model.
The horizontal lines represent time lines of the links.
The shaded areas represent the only burst of activity of each link, where $w$ interaction events (vertical lines) take place during a time period of length $\Delta$.
This pattern of activity repeats periodically in time with period $T$.
}
\label{fig:3}
\end{figure}

Next, we will attempt a simple explanation for the  linear relationship between $\eta$ and $\mu^{H}$. To this end, we introduce a simple temporally heterogeneous network model that can be analytically addressed (see Fig.~\ref{fig:3}).
In the Single-Burst (SB) model, each link has only a single burst of activity during a time period of duration $T$. (Note that we still apply periodic boundary conditions.) This single burst spans a time interval of $\Delta\leq T$, and during the burst, $w$ events take place. The bursts of different links are independent of each other.

Our aim is to find an expression relating $\eta$ and $\mu^{H}$ for the the model. 
To estimate $\eta$, consider how spreading dynamics unfold on top of the model.
Initial infection occurs at time $t_0$ at node $v_0$. Then,
eventually, the infection reaches the $\frac{N}{2}$-th infected node through a chain of $\ell$ infection events occurring at times $t_1 < t_2 < \dots < t_{\ell}=t_{1/2}$. 
The time differences $\tau_R(j)=t_j-t_{j-1}$, for $j=1,...,\ell$ are the waiting times (\emph{relay} times) at the links along the chain~\cite{kivela2012multiscale}.
In essence, after the infection reaches node $v_{j-1}$ through link $e_{j-1}$ at time $t_{j-1}$, it has to wait until the next contact occurring at time $t_j$ via another link $e_j\neq e_{j-1}$, infects the next node $v_j\neq v_{j-1}$ in the chain. 
We can then roughly estimate that on average $\avrg{t_{1/2}}=\avrg{\sum_{j=1}^{\ell} \tau_R(j)}\simeq \avrg{\ell}\avrg{\tau_R}$.
The average relay time at the links, $\avrg{\tau_R}$, can be obtained by using order statistics~\cite{david2003order},
\begin{equation}
\label{eq:4}
\avrg{\tau_R} = \frac{T}{w+1}\left[1+\frac{w-1}{2}\left(1-\frac{\Delta}{T}\right)^2\right].
\end{equation}
For the UT reference model where there are no temporal heterogeneities, i.e.\ the bursts have been replaced by uniformly distributed events,  we similarly have $\avrg{t_{1/2}}_{\mathrm{UT}}\simeq \avrg{\ell}_{\mathrm{UT}}\avrg{\tau_R}_{\mathrm{UT}}$, and now the average relay time $\avrg{\tau_R}_{\mathrm{UT}}=T/(w+1)$ as can be seen by setting $\Delta=T$ in Eq.~(\ref{eq:4}).
Now $\eta=\left({\avrg{t_{1/2}}_{\mathrm{UT}}}\right)/\left({\avrg{t_{1/2}}}\right)\simeq 
\left(\avrg{\ell}_{\mathrm{UT}} \avrg{\tau_R}_{\mathrm{UT}}\right)/\left(\avrg{\ell} \avrg{\tau_R}\right)$.
Therefore, if we assume that $\avrg{\ell}_{\mathrm{UT}} \simeq \avrg{\ell}$, we may approximate 
\begin{equation}
\eta\approx \frac{\avrg{\tau_R}_{\mathrm{UT}}}{\avrg{\tau_R}}.
\label{eq:5}
\end{equation}
Using Eq.~(\ref{eq:4}), we obtain
for the SB model 
\begin{equation}
\label{eq:6}
\eta \simeq  \left[1+\frac{w-1}{2}\left(1-\frac{\Delta}{T}\right)^2\right]^{-1}.
\end{equation}

\begin{figure}
  \includegraphics[width=.65\columnwidth]{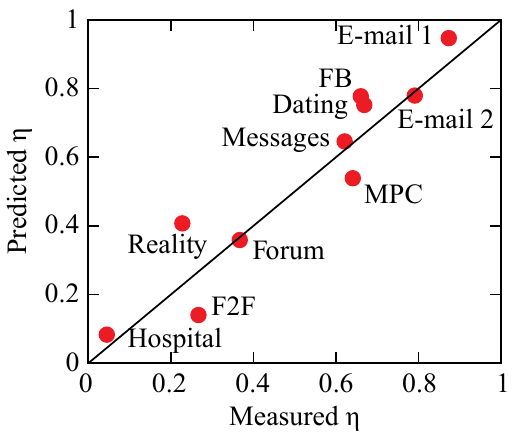}
    \caption{
(Color online).
Comparison between values of $\eta$ predicted using the estimation from the SB model, Eq.~(\ref{eq:7}), against the measured values of $\eta$ from numerical simulations over the temporal network data sets (red circles and labels). Predicted $\eta$ follows measured $\eta$ almost linearly, with $R^2=0.88$ between the data points and identity function (solid line).
}
\label{fig:4}
\end{figure}

To estimate the sparsity coefficient $\mu^{H}$, consider the probability $p_e$ that one of the events of link $e$ takes place in a given time window.
We approximate this probability as $p_e \simeq \Delta/T \simeq \mu^{H}$.
Therefore we obtain the expression relating $\eta$ and $\mu^{H}$ for the SB model by rewriting  Eq.~(\ref{eq:6}) as
\begin{equation}
\label{eq:7}
\eta \simeq
\left[1+\frac{w-1}{2}\left(1-\mu^{H}\right)^2\right]^{-1}.
\end{equation}
This equation clearly points out that temporal sparsity slows down spreading. However, it is obvious that we do not recover a simple linear relationship between $\eta$ and $\mu^{H}$ for the model (also note that $w$ and $\mu^{H}$ are not independent). Despite this,
at its limits, Eq.~(\ref{eq:7}) behaves as it should. If a network has no temporal heterogeneities, $\mu^{H}\simeq 1$, and it follows that $\eta \simeq 1$, as expected.
On the other hand, a highly temporally heterogeneous network corresponds to $\mu^{H}\ll 1$, and thus $\eta \simeq \frac{2}{w+1}$.
For $w\simeq 1$ this quantity also converges to one, which is correct, as a burst of activity with only one event is not meaningful.
For $w\gg 1$ we get $\eta \simeq 2/w \ll 1$. In other words, there is a strong slowing-down effect, as expected. 

Why does the empirical relationship between $\eta$ and $\mu^H$ appear linear (Fig.~\ref{fig:2}), while Eq.~(\ref{eq:7}) shows a more complex dependence? It is evident that real-world temporal networks are different from the simple SB model, and this may give rise to a different dependence between $\eta$ and $\mu^H$. In real-world networks, there are numerous bursts per link, and the activity levels of nodes and links are typically broadly distributed, as is the number of events in each burst~\cite{Karsai2012a}.

Despite the simplicity of the model, it is interesting to see how the model works with real data.
To this end, we have computed
$\eta_\mathrm{model}$ by estimating the model parameters for Eq.~(\ref{eq:7}) from the empirical networks, inserting the values $w=E/M$ and $\mu^{H}$ computed for each of the data sets. The results match the empirically observed values of $\eta$ remarkably well (Fig.~\ref{fig:4}). This is surprising considering the simplicity of the single-burst model. However, it highlights the importance of the temporal sparsity coefficient $\mu^H$, which is the only connection between the empirical networks and the model that carries information about the level of temporal heterogeneities.

The approximations made in deriving Eq.~(\ref{eq:7}) are worth considering. Here, a key approximation is that spreading proceeds along paths where the burstiness-induced waiting times neatly sum up, leading to Eq.~(\ref{eq:5}). This approximation does not hold in general (e.g.~in our empirical networks); however it works for the SB model given a proper range of parameter values. 
Numerical simulations of the SB model on top of $k$-regular graphs indicate that Eq.~(\ref{eq:5}) fails when the \emph{effective degree} $k_\mathrm{eff}^{H}=2 M_\mathrm{eff}^{H}/N\gg 1$. For $k_\mathrm{eff}^{H} \lesssim 1$, waiting times at the nodes are of the order of $\tau_R$, and Eq.~(\ref{eq:5}) yields the correct approximation. In this regime, the spreading process has only a single or a few ways forward from a node in a given time window, and on average has to wait for $\tau_R$ time units at each link to proceed. Note that our empirical networks are typically in this regime with $k_\mathrm{eff}^{H} \lesssim 1$ (see Table~\ref{table:table1}). On the other hand, if the number of available links is large and the network is temporally dense, the average waiting time to the activation of the \emph{first} of these links is smaller than $\tau_R$~\cite{inpreparation}.

In this work, we have shown  that temporal heterogeneities make temporal networks sparser at the natural time scale of average inter-event times.
This temporal sparsity makes the Susceptible-Infectious spreading process slower, with a linear dependence between the level of slowdown and the level of temporal sparsity for a number of empirical temporal networks of different origins.
Because the SI process always follows the fastest temporal paths, our finding generalises to other spreading processes as well. The more temporally sparse a network is, the longer it takes for anything to be transmitted between pairs of nodes; in other words, temporal sparsity increases the latency of temporal paths. In addition to the empirical data, we confirm this finding with an approximate analytical solution for a simple temporal network model with a single burst of events at each link.

\begin{acknowledgments}
J.I.P. and J.S. acknowledge support by the Academy of Finland, project No.\ 260427.
H.-H.J. was supported by the Aalto University postdoctoral program.
P.H. was supported by the Basic Science Research Program through the National Research Foundation of Korea (NRF) funded by the Ministry of Education (2013R1A1A2011947).
\end{acknowledgments}

\end{document}